\documentclass[a4paper,fleqn]{cas-sc}

\usepackage[numbers]{natbib}

\def\tsc#1{\csdef{#1}{\textsc{\lowercase{#1}}\xspace}}
\tsc{MK}
\tsc{MRA}

\begin{document}
\let\WriteBookmarks\relax
\def\floatpagepagefraction{1}
\def\textpagefraction{.001}
\shorttitle{GenFloor: Interactive Generative Space Layout System via Encoded Tree Graphs}
\shortauthors{Keshavarzi and Rahmani-Asl}

\title [mode = title]{GenFloor: Interactive Generative Space Layout System via Encoded Tree Graphs}

\author[1]{Mohammad Keshavarzi}[
                        orcid=0000-0003-2881-165X]
\cormark[1]
\ead{mkeshavarzi@berkeley.edu}


\author[2]{Mohammad Rahmani-Asl}[
   ]
\ead{mohammad.asl@autodesk.com}


\address[1]{Department of Architecture, University of California, Berkeley, CA, USA}
\address[2]{Autodesk Inc., San Francisco, CA, USA}

\cortext[cor1]{Corresponding author}

\begin{abstract}
Automated floorplanning or space layout planning has been a long-standing NP-hard problem in the field of computer-aided design, with applications in integrated circuits, architecture, urbanism, and operational research. In this paper, we introduce GenFloor, an interactive design system that takes geometrical, topological, and performance goals and constraints as input and provides optimized spatial design solutions as output. As part of our work, we propose three novel permutation methods for existing space layout graph representations, namely O-Tree and B*-Tree representations. We implement our proposed floorplanning methods as a  package for Dynamo, a visual programming tool, with a custom GUI and additional evaluation functionalities to facilitate designers in their generative design workflow. Furthermore, we illustrate the performance of GenFloor in two sets of case-study experiments for residential floorplanning tasks by (a) measuring the ability of the proposed system to find a known optimal solution, and (b) observing how the system can generate diverse floorplans while addressing given a constant residential design problem. Our results indicate convergence to the global optimum is achieved while offering a diverse set of solutions of a residential floorplan corresponding to local optimums of the solution landscape. 
\end{abstract}


\begin{keywords}
Generative Design\sep Floorplanning\sep Optimization\sep Space Layout\sep  Performance-based Design\sep Optioneering
\end{keywords}

\maketitle

\section{Introduction}
Space layout planning is the process of placing a set of discrete but independent spatial elements while attempting to satisfy geometrical, topological, and performance goals in their layout. Due to the combinatorial enumeration of spatial relationships and the subjective nature of their inter-dependencies, large numbers of potential solutions can be generated even with a small number of space elements. Such complexity classifies the floorplanning process as an NP-Hard problem, where the size of solution spaces grows exponentially as the size of the problem increases \cite{jo1998space}. Applications of floorplanning can be seen in a wide variety of fields, such as integrated circuits \cite{Singh2016}, architecture, urbanism, and operational research \cite{Anjos2017}. More recently, with the rise of augmented and virtual reality platforms, the need to automate surrounding object placements has surpassed virtual applications \cite{keshavarzi2020optimization}, and tasks such scene synthesis \cite{zhang2019survey} and indoor object augmentation \cite{keshavarzi2020scenegen} take advantage of various implicit or explicit space layout methodologies, and are seen as promising research directions for next-generation computing platforms.

This level complexity of floorplanning has encouraged researchers to explore this problem by developing generative systems that take advantage of meta-heuristic search approaches. The general process of such systems can be described in three modules (i) generating various layouts via a generative function (ii) analyzing the layouts using certain design objectives via a fitness function, or comparing them to a set of priors; (iii) iterating this search until the optimal solutions are found. Generating layouts often involves encoding a floorplan as a solution to a floorplan representation function. This representation not only defines the complexity of the solution but also impacts the efficiency of the search process to find a desired floorplan in the solution landscape. Evaluation metrics can vary based on the design problem itself. In architectural space planning, for instance, performance metrics such as functional topologies, daylight, energy consumption, etc. come along with geometrical requirements such as site boundaries, room areas, etc. In digital design and VLSI (very large scale integrated-circuit) floorplanning, factors such as wire length and bounding area are optimized while fulfilling the requirements of the fabrication technology (sufficient wire spacing, restricted number of wiring layers, etc.). As each performance objective induces a new dimension within the solution space, searching for the desired solution has been widely incorporated with meta-heuristic search methods such as Genetic Algorithms, Simulated Annealing, Particle Swarm Optimizations, Ant Colony Optimization, etc.  

While much of the floorplanning methodologies proposed in current literature can be incorporated into cross-disciplinary challenges, many current space layout systems only target a specific range of design applications. Floorplanning software used by VLSI designers, for instance, is not often used by architects, and vice versa. This is mainly due to fact that such systems differ in performance evaluation criteria and are designed primarily for users targeting a specific application. Moreover, choosing the best heuristic optimization method can also be dependent on the design problem itself. Previous studies have compared these search methods in various floorplanning benchmarks, showing how each may perform better than the other in different design requirements \cite{Youssef2001,Caldas2001,Singh2016}. However, not all optimization methods can be utilized in current floorplanning systems for a non-expert user to explore different optimization strategies. Such investigation, that would incorporate custom performance analysis and allow the user to choose between different search methods often requires advanced proficiency in programming, while allocating considerable time and resources. 

In the past two decades, many designers have adopted visual-programming languages (VPLs) such as Grasshopper and Dynamo to automate their design workflows. VPLs break down the functions of software into discrete \textit{components} or \textit{nodes} using a intuitive graphical interface. Each component within a VPL has its own inputs and outputs, essentially acting as an individual tool within a larger script or workflow. As a result, users can customize their workflows based on the arrangement of components in their scripts, enabling them to engage different issues as they become relevant and experiment with new creative workflows as unique situations arise on projects \cite{tedeschi2014aad}. Moreover, many building performance evaluation libraries and third-party tools which facilitate bi-directional data-flow between simulation engines have become increasingly available through these platforms. In addition, meta-heuristic optimization libraries such as SPEA-2 \cite{Zitzler2001}, HypE \cite{bader2011hype}, NSGA-2 \cite{deb2000fast} have been integrated in various VPL plugins, allowing users to perform generative design task, and visualize the results using the incorporated visualization tools.  However, with all the design opportunities VPLs have brought for generative designers, there are a limited number of end-to-end workflows available in VPLs that take advantage of established cross-disciplinary floorplanning representations, which are specifically designed for the meta-heuristic optimizations available in VPLs.

Motivated by this challenge, in this paper we introduce an interactive generative floorplanning design system, GenFloor, which is developed as a library package for Autodesk Dynamo, allowing users to incorporate generative floorplanning mechanism with third-party performance evaluators for optimization purposes within commonly used CAD software. With a user-centric design approach and an end-to-end workflow, GenFloor takes geometrical, topological, and performance goals and constraints as input and provides optimized spatial design solutions as output. By proposing three novel permutation methodologies for binary and non-binary trees, and integrating them with meta-heuristic optimization algorithms and defined performance evaluators, GenFloor enables designers to generate and optimize Manhattan-based floorplans for various design applications.

\section{Related Work}
Since the early 1960s, various methods have been developed for the computer-based solution of floorplanning problems \cite{eldars1964approach,frew1980survey}. One of the earliest work is of Armor and Buffa's in {\small{CRAFT}} \cite{armour1963heuristic}. {\small{CRAFT}} executed automated floorplanning tasks on Sketchpad, the first CAD system developed by Sutherland. Floorplanning research followed with the work of of Grason \cite{grason1970fundamental},  Miller \cite{miller1971computer},  Ligget \cite{liggett1981optimal} and  Mitchel \cite{mitchell1975theoretical} in the next two decades. Work of March and Steadman \cite{march1971spatial} and Shaviv \cite{shaviv1987generative} review the first 30 years of spatial allocation algorithms.

The method that was employed in nearly all floorplanning and space layout systems at the time was a combination of a generative mechanism for the creation of alternative variations and an evaluation mechanism to assess these variations \cite{mitchell1990logic}. However, many classical approaches attempted to exhaustively search through all possible arrangements with a specified number of rooms. In the 1990s, Flemming presented systems for generating layouts using constraint programming which was developed under the name {\small{SEED}} \cite{flemming1995software} and {\small{LOOS}} \cite{flemming1990planning, flemming1989more}. LOOS used orthogonal structures for the representation of loosely packed arrangements of rectangles. In the same decade, early examples of the use of evolutionary approaches for floorplanning in architecture were explored by Tam \cite{tam1992genetic} and Gero and colleagues \cite{jo1995representation,gero1997learning,schnier1996learning,jo1998space,rosenman1999evolving}. An example of such an approach can be seen in Jo and Gero's work \cite{jo1998space} where topological and geometrical goals of spatial elements were encoded in a GA for a large office layout problem. Though not many variations of the spatial arrangement were highlighted, their work showed the robustness of coupling genetic algorithm-based searches with design workflows to produce good results in space planning.

In the 2000's Work of Michalek, et al. \cite{Michalek_Choudhary_Papalambros_2002}, Knecht \cite{knecht2010generating}, and Doulgerakis \cite{doulgerakis2007genetic} took advantage of graph and tree-based representation previously used in VLSI floorplanning \cite{kahng2011vlsi,Wang:2009:EDA:2843514} and were utilized with various optimization search methods in architectural contexts. Arvin and House \cite{arvin2002modeling} apply physically-based modeling to layout optimization, representing rooms and adjacencies as a mass-spring system. This heuristic has only been demonstrated on collections of rectangles and is sensitive to the initial conditions of the system. Nassar \cite{Nassar_2010} presented new findings in graph theory with direct implications in space planning problems. He described architectural space plans as simple, connected, labeled planar graphs, and elaborated on the relevance of finding a rectangle dual for every planar graph to increase solution space. This work outlined a tool for architects to generate space plans. Realizing spatial relationships as planar graphs with nodes as rooms and edges as adjacencies, it claimed that the proposed model could provide a truly exhaustive set of potential designs. Work of Slusarczyk et al.\cite{slusarczyk2018graph} follows a similar approach, where they propose a framework for supporting the design process by defining design requirements over Hierarchical layout hypergraphs (HL-graphs) \cite{grabska2007hierarchical} representations of floorplan designs.

More recently, researchers in the realm of Machine Learning have explored how Generative Advisoral Networks (GAN) can facilitate the space layout problem in architectural contexts. Originally introduced by Goodfellow et al, \cite{goodfellow2014generative} GANs leverage a feedback loop between a Generator and Discriminator model, to slowly build an ability to create synthetic data, factoring in phenomena found among observed data. The work of Isola et al \cite{isola2017image} in Pix2Pix extends this generative ability to images, by allowing networks to learn a proper mapping from one image to another. Huang and Zheng \cite{huang2018architectural} take advantage of Pix2PixHD \cite{wang2018high} to recognize and generate furnished architectural drawings. They do this by translating floorplan images to programmatic patches of color, and inversely, generate patches of color that are turned into floorplans.  Chaillou expands this approach by nesting three models (footprint, program, and furnishing) to generate floorplans, given a set of initial conditions and constraints. Extracting and manipulating existing layouts from images and sketches using model retrieval methods \cite{keshavarzi2020sketchopt, bergin2020automated} has also been used as a data augmentation method for generating new layouts \cite{keshavarzi2020genscan}. 

However, there are two main drawbacks in current GAN floorplanning methods which can be highlighted especially in performance-oriented design scenarios. First, as with any machine-learning model, GANs learn from statistically significant features from the input data. Hence, the discriminator’s ability to classify various design generations is dependent on the data initially inputted to the discriminator. Given the fact that the reviewed systems above take advantage of the Pix2Pix model, the significant features required for the learning model are extracted from visual features of input floorplan pictures. Therefore, in performance-based design scenarios, the user must ideally provide images of pre-designed layouts with similar design constraints and performance goals, which may be difficult to gather subject to the nature of the problem. Even from a visual standpoint, the results generated by GAN floorplanning systems may lack emergence and layout diversity when limited to previously designed datasets. Second, training current GAN systems is time-consuming procedures and often achieved with high-performance computational hardware. If we consider the design process as an iterative process, where the designer may constantly change its performance priorities and goals based on various outputs in each design step, re-training and running such systems to alter design strategies may be a time-consuming procedure.

Floorplanning is also considered an essential step in the VLSI physical design process. In such context, floorplanning methodologies mostly aim at optimizing chip area and wire lengths to reduce interconnections and improve timing \cite{kahng2011vlsi}. The floorplan representation plays an important role in the robustness of the algorithm \cite{jain2013non} and determines the size of the search space and the complexity of transformation between a representation and its corresponding floorplan. Floorplanning representations are generally divided into two main categories: slicing and non-slicing representations \cite{Wang:2009:EDA:2843514}. In slicing methodologies, the floorplan is recursively bisected until each part consists of a single module. Commonly represented by a binary tree, modules are represented at the leaves of the tree and horizontal and vertical cuts are represented as internal nodes. The normalized Polish Expression \cite{Wong:1986:NAF:318013.318030} is considered as a popular slicing representation, where the post-fix ordering of a binary tree is implemented after post-order traversal on a binary tree and does include any node of the same cut type as its right child.

Non-slicing representation are utilized for more general use cases where no recursive bisection of a certain area takes place.\cite{nakatake1997module, Chang_Chang_Wu_Wu_2000,gwee1999ga,Guo_Cheng_Yoshimura_2003, ma2001vlsi,lin2005tcg} Such representation are often defined in constraint graphs defining the horizontal and vertical relations among the modules. While no bisection of specific boundary is performed, non-slicing representations can be bounded by a given fixed outline.  Representations such as Sequence Pair \cite{Murata:1995:RMP:224841.225094}, Bounding Slicing Grid \cite{nakatake1997module}, B*Tree \cite{Chang_Chang_Wu_Wu_2000}, Integer Coding \cite{gwee1999ga} , OTree\cite{Guo_Cheng_Yoshimura_2003}, Corner Block List\cite{ma2001vlsi},Transitive Closure Graph \cite{lin2005tcg} are among widely used non-slicing methodologies. Multiple studies have integrated these representations with various optimization algorithms such as Simulated Annealing (SA) \cite{kirkpatrick1983optimization, kiyota2005simulated, Wong:1986:NAF:318013.318030}, Genetic Algorithms (GA) \cite{rebaudengo1996gallo,nakaya2000adaptive, lin2002efficient,gwee1999ga,xiaogang2002vlsi} and Particle Swarm Optimization (PSO) \cite{sun2006floorplanning,chen2008vlsi,kaur2014enhanced,sowmya2013minimization,moni2009vlsi}. These methods are also used in different versions of commercial floorplanning software such as PARQUET Floorplanner\cite{adyaparquet} where an inbuilt SA optimizer is implemented for the search and optimization process. Our system takes advantage of non-slicing OTree and B*Tree representations but differs in the perturbation methodology previously used in these methods. 

It is important to note, all representations covered above provide an initial general layout for spatial elements. Additional operators can be implemented to transform the rectangular solutions into more complex geometries, or allocated circulations space in applications such as architecture and facility layout planning. Such an approach would result in more satisfactory results. An example of such an approach can be seen in the work of Marson and Musse \cite{Marson_Musse_2010} for instance, implemented a squarified treemap algorithm \cite{bruls2000squarified} to compartmentalize the input space into different zones or regions. These zones are organized into a hierarchy that satisfies design goals and site constraints and is visualized into a square treemap to generate various floor layouts. Their work excels at building sophisticated circulation networks. After the rooms are located in the boundary, a circulation network graph is built to understand which rooms are connected or disconnected from each other. They use an A* algorithm to traverse the connectivity graph and find the shortest paths to access spaces from the lobby. 

There is also a large body of literature focused on how to interpret generative design solutions, and to provide the user with appropriate workflows to modify and interact with the generated solution space \cite{scott2002investigating,natureHardy,meignan2015review}. Chaszar et al. \cite{Chaszar2016MultivariateIV} explored methods and tools for multivariate interactive data visualization of the generated designs and simulation results by enabling designers to not only focus on high-performing results but also examine suboptimal ones. Mueller and Ochsendorf \cite{mueller2015combining}  proposed a computational design exploration approach that takes advantage of an interactive evolutionary algorithm to integrate the designers' preferences within the solution search. Work of Keshavarzi et al. \cite{keshavarzi2020v} extends the user interaction with the solution space from 2D input to 3D exploration, by developing a virtual reality generative analysis framework for navigating large-scale solution spaces. In this work, we take advantage of native generative design visualization and exploration tools in Dynamo VPL, enabling the user to visualize, filter and analyze Pareto-optimal solutions in real-time in different performance metrics.

The work of Das et al. \cite{das2016space} is probably the most similar work to our system where they also take advantage of efficient tree representation via Kd-trees as an initial generation method. Furthermore, they implement spatial grouping and circulation functions to achieve more feasible architectural floorplanning results. Their system is also developed as a library that can be loaded to Dynamo VPL, allowing users to generate and evaluate various floorplanning representations. However, from our analysis of their system, their generative model includes multiple random functions and therefore does not generate deterministic results on multiple runs. Such attributes can cause challenges for solution convergence in evolutionary solvers, especially population-based meta-heuristic optimization algorithms, where different results are generated in each generation of the search. In contrast, our system uses a deterministic encoding mechanism, providing a constant result for the same input and does not integrate any random functions with its generative module.

 \begin{figure*}
  \centering
  \includegraphics[width=1\columnwidth]{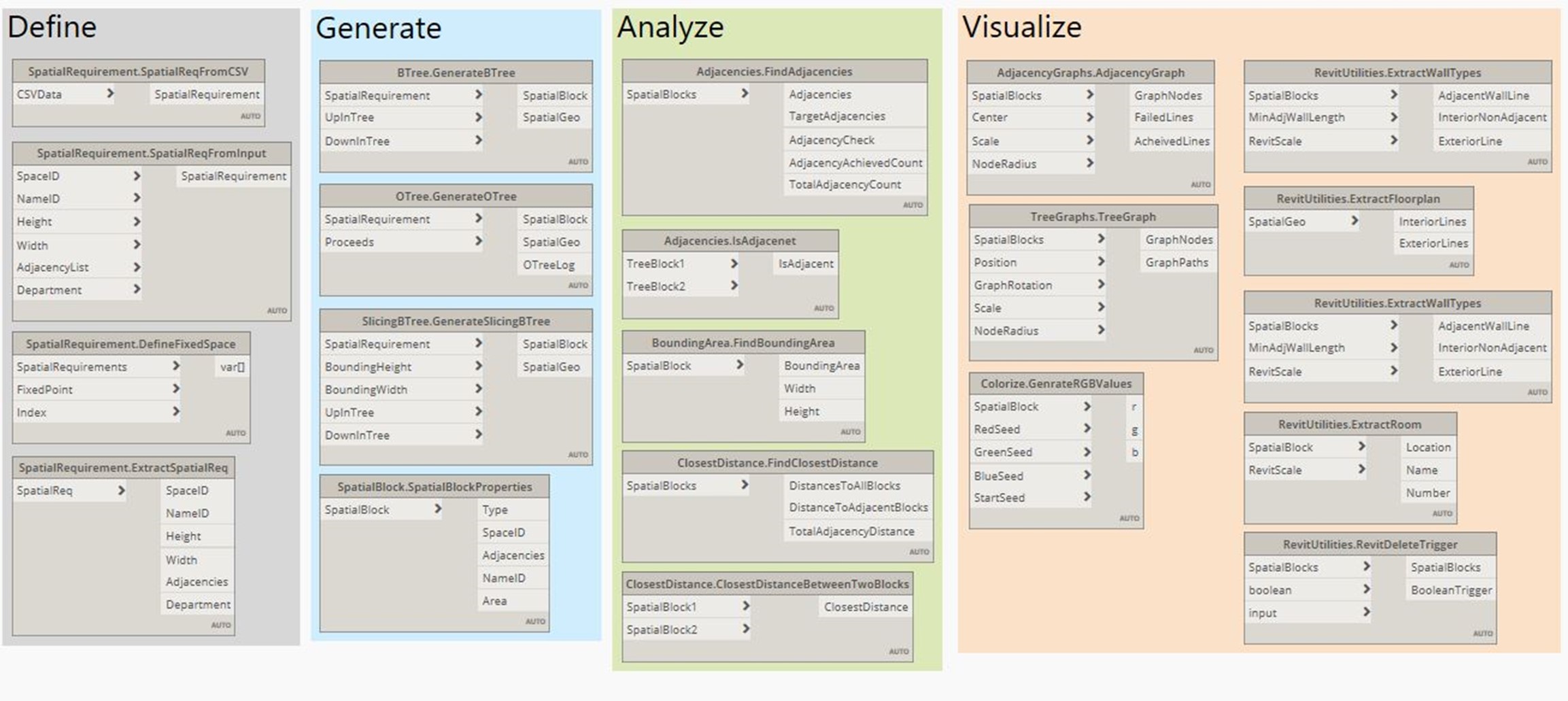}
  \caption{GenFloor components available in Dynamo VPL. The Define components provide SpatialRequirements for the Generate module, which outputs SpatialBlocks. The SpatialBlocks are further analyzed and use as fitness for the search mechanism. Finally, the outcome of the results can be visualized using the corresponding components.
  \label{fig:all_comp}}
\end{figure*}

\section{Methodology}
GenFloor is developed with the goal to be utilized by users with limited proficiency in visual scripting and programming while allowing advanced users to be able to modify and adjust input parameters of the system, dependent on their needs. As a first step, users define the geometrical and topological requirements of each space by producing \textit{Spatial Requirement} entities. Next, using encoding functions, a tree structure and its corresponding \textit{Spatial Blocks} are generated, resembling a spatial layout corresponding to an array of \textit{Permutation Parameters}. The Spatial Blocks can then be analyzed using various performance evaluators customized by the user. Finally, by integrating the system with a meta-heuristic multi-objective search mechanism, the optimization of the floorplans takes place allowing the system to find optimal tree permutations and subsequently their corresponding floorplans matching the defined goals and constraints specified by the user.

We implement our proposed methodology as a software package for the Autodesk Dynamo platform. Dynamo is a visual programming language (VPL) used by engineers, architects, and urban designers allowing custom visual scripting in conventional design and engineering software such as Revit, Fusion, and FormIt. Moreover, visual programming platforms allow third-party developers to build various add-on libraries and custom functions as packages. Such packages can serve as custom performance-based analysis functions when integrated with GenFloor’s generative design framework. Floorplans generated via GenFloor can be analyzed using custom libraries available in the visual programming platform developed by other developers, while in contrast, standalone software outside VPL platforms may not have the ability of such integration. GenFloor is developed using \textit{ZeroTouchNodes} and \textit{CustomUINodes} libraries which are all written in C\# language.

In the following subsections, we elaborate on the details of the system in four major steps, commonly used in a generative design workflow: (1) Define, where the user provides the system the Spatial Requirements; (2) Generate, where the system generates the Spatial Blocks; (3) Evaluate and Search, where the fitness of each solution is calculated and is input to the ongoing search (4) Visualize, where the final floorplan, it's tree representation and adjacency graphs are visualized or converted to a BIM model.

\section{Define}
\begin{figure}
  \centering
  \includegraphics[width=1\columnwidth]{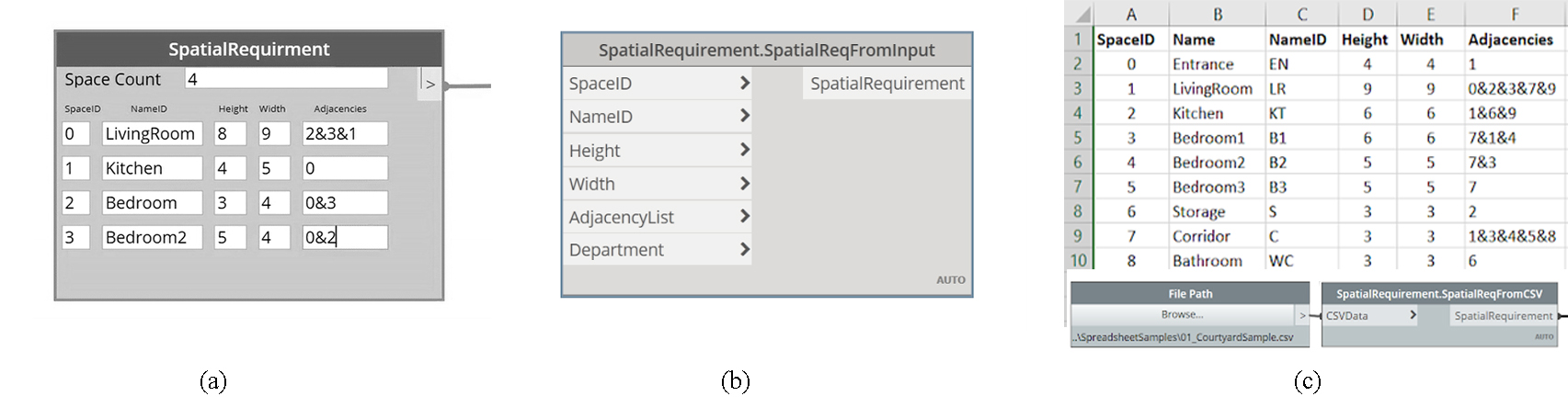}
  \caption{Various types of Define nodes in the GenFloor library for generating the SpatialRequirement data (a) Custom GUI node in the Dynamo interface where users can manually input the data (b) data can be imported from other data streams and nodes (c) data can be imported from an external CSV file.
  \label{fig:define}}
\end{figure}

As a starting step, users of the system define geometrical and topological spatial requirements by either manually defining each spatial block using a customized interactive GUI node in the Dynamo interface, input from other dynamic data streams in Dynamo, or importing program information from an excel spreadsheet. All three approaches are shown in Figure \ref{fig:define}. Block dimensions and adjacency goals can be defined in this step. Defining the spatial requirements using dynamic data streams would allow more advanced users to parametrize the input data with more control (eg. parameterize the dimension of the block), while the spreadsheet data parsing method is targeted for general and traditional floorplanners that organize their program requirements in spreadsheet format. The data inputted in this step is primarily used for generating the tree structures but can also serve as parameters defining the floorplanning goals and constraints.

\begin{figure}
  \centering
  \includegraphics[width=0.6\columnwidth]{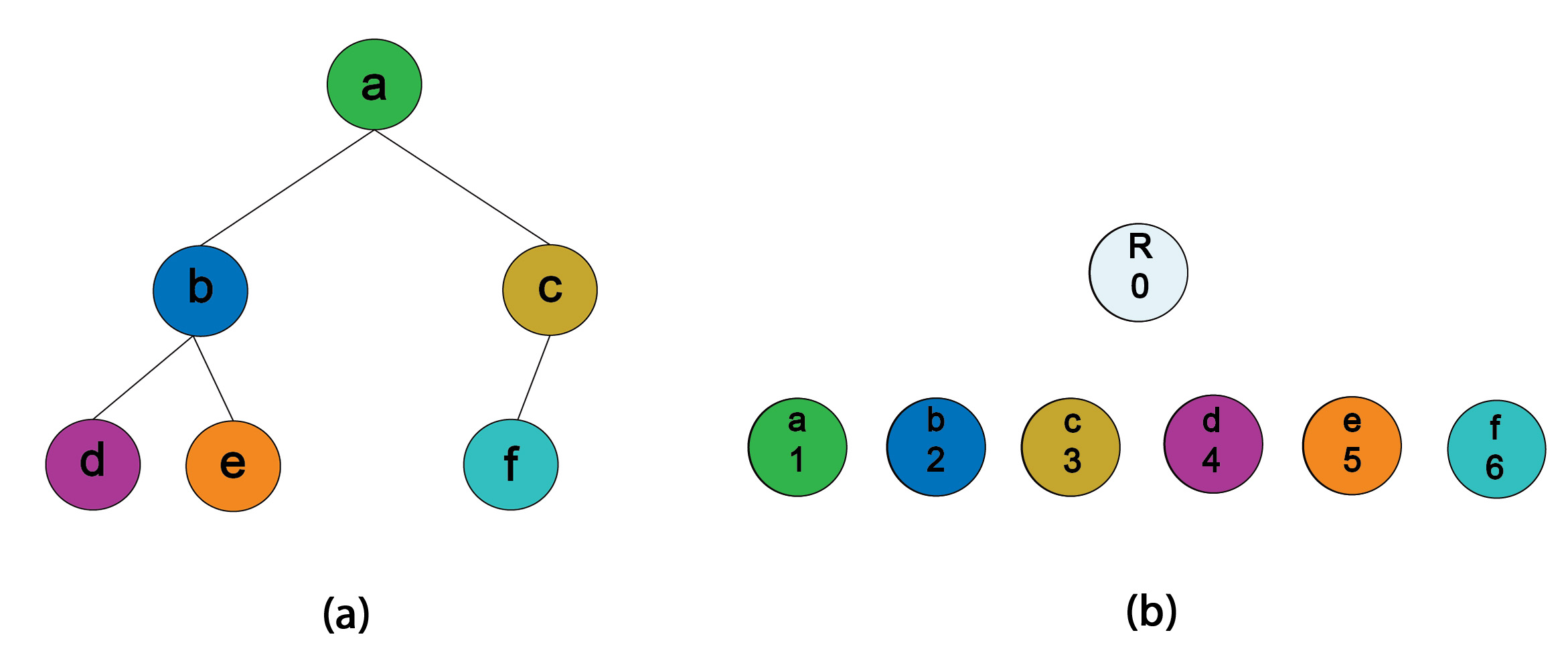}
  \caption{Standard Tree of a (a) binary tree for B*Tree and (b) non binary tree for O-Tree. $n = 6$ and each node corresponds to an spatial input entity (a-f).
  \label{fig:standardTree}}
\end{figure}

 By knowing the quantity of the blocks $(n)$, a \textit{Standard Tree} with $n$ nodes is generated. Since we use an O-Tree and B*Tree representation in our system, we develop two methods of Standard Tree generation for each representation. In a B*Tree generation scenario, a Standard Tree is a complete binary tree of $n$ nodes, where each node corresponds to Spatial Requirement defined by the user. A complete binary tree is a binary tree in which every level, except possibly the last, is filled, and all nodes are as far left as possible. The order of the nodes in the Standard Tree resembles the order in which the user inputted each entity of the geometrical block properties. Therefore, the first entity corresponds to the tree root and the final entity is the far-right node on the last level. In an O-Tree scenario, the nodes are located on one single level, ordered from left to right. Figure \ref{fig:standardTree} shows an example of a Standard Tree for a B*Tree and O-Tree.

 \section{Generate}
 
 The goal of this module is to convert the Spatial Requirements to a tree representation and further to a set of Spatial Blocks using the Permutation Parameters. As part of our work, we propose three novel permutation methods for existing space layout graph representations, namely O-Tree and B*-Tree representations to perturb the Standard Tree with Permutation Parameters. Permutation Parameters $P$ are an array of values, which vary in size and attributes depending on the permutation method. We elaborate more on the properties of different types of Permutation Parameters when describing each permutation method. Our methods provide deterministic results of tree arrangements and do not utilize any random functions in their process. After perturbing the Standard Tree with one of the permutation modules, the subsequent tree is converted to a floorplan using O-Tree or B*-Tree representations. Users of our system can generate space layouts by choosing one of the block generation methods. In simple terms, the Generation modules take Spatial Requirement data and Perturbation Parameters as input and output a list of Spatial Blocks which formulate together a determined layout.
 
 To facilitate the explanation of our system, we initially describe how O-Tree and B*Tree non-slicing floorplanning representations are defined, and then present our proposed permutation methods. Furthermore, we explain our implementation with the goal of allowing non-expert users to generatively design and manipulate floorplanning outputs. Finally, we introduce a procedural mechanism to extend the non-slicing representations to a slicing representation, allowing our system to generate layouts for applications such as indoor furniture layout.

 \subsubsection{O-Tree Representation}\label{sec:otree}
 An O-tree is an ordered non-binary tree representing a determined floorplan layout. We define and implement a modified version of O-Trees first introduced by Guo et. al \cite{Guo_Cheng_Yoshimura_2003} mainly altering perturbation methodologies. Given a non-binary tree of function blocks, we first order the nodes using a depth-first search (DFS). Note that the root node does not correspond to any function block and represents the left boundary of the placement. To generate the floorplan using the ordered nodes we take the following steps for each function block. Each child $j$ of a parent node $i$ is placed on the right side of the parent node $i$ where $ x_i =  x_j + w_j $. For determining the height of the function block placement, the block is placed on top of the horizontal contour of the placement. The horizontal contour on the top edge of the previously placed function blocks the floor planning placement and is updated recursively when each function block is placed.
 
  \subsubsection{B*Tree Representation}\label{sec:btree}
 A B*-Tree is an ordered binary tree representing a floorplan layout. Our implementation of the B*-Tree floor planning representation is a modified version of B*-Tree\textsc{\char13}s initially introduced in Chang et al study \cite{Chang_Chang_Wu_Wu_2000} while having major differences in the perturbation operations. In a B*Tree representation, the block assigned to the root node is placed on the bottom-left corner. Using a depth-first search (DFS) procedure, we recursively construct the floor-plan layout by keeping the geometric relationship between two modules as follows. If node $b$ is the left child of node $a$, the block assigned to node $b$ must be located on the right-hand side to the block assigned to node $a$ in the floor planning placement such as $ x_b =  x_a + w_a $; where $x_i$ is the starting point and $w_i$ is the width of block $i$. During the placement of the function blocks, we also consider assigning the height of the function block based on the horizontal contour, which we explained in the previous section. Furthermore, if node $c$ is the right child of node $a$, the block assigned to $c$ must be located above the horizontal contour and also the block assigned to block $a$. As the construction of the floorplan layout is recursively performed in the DFS order,  we first complete the left subtree of each node and then apply the same procedure to the right subtree of the node.

 \subsection{Proceeding Perturbation of O-Tree Representations}\label{sec:procper}

\begin{figure*}
  \centering
  \includegraphics[width=1\columnwidth]{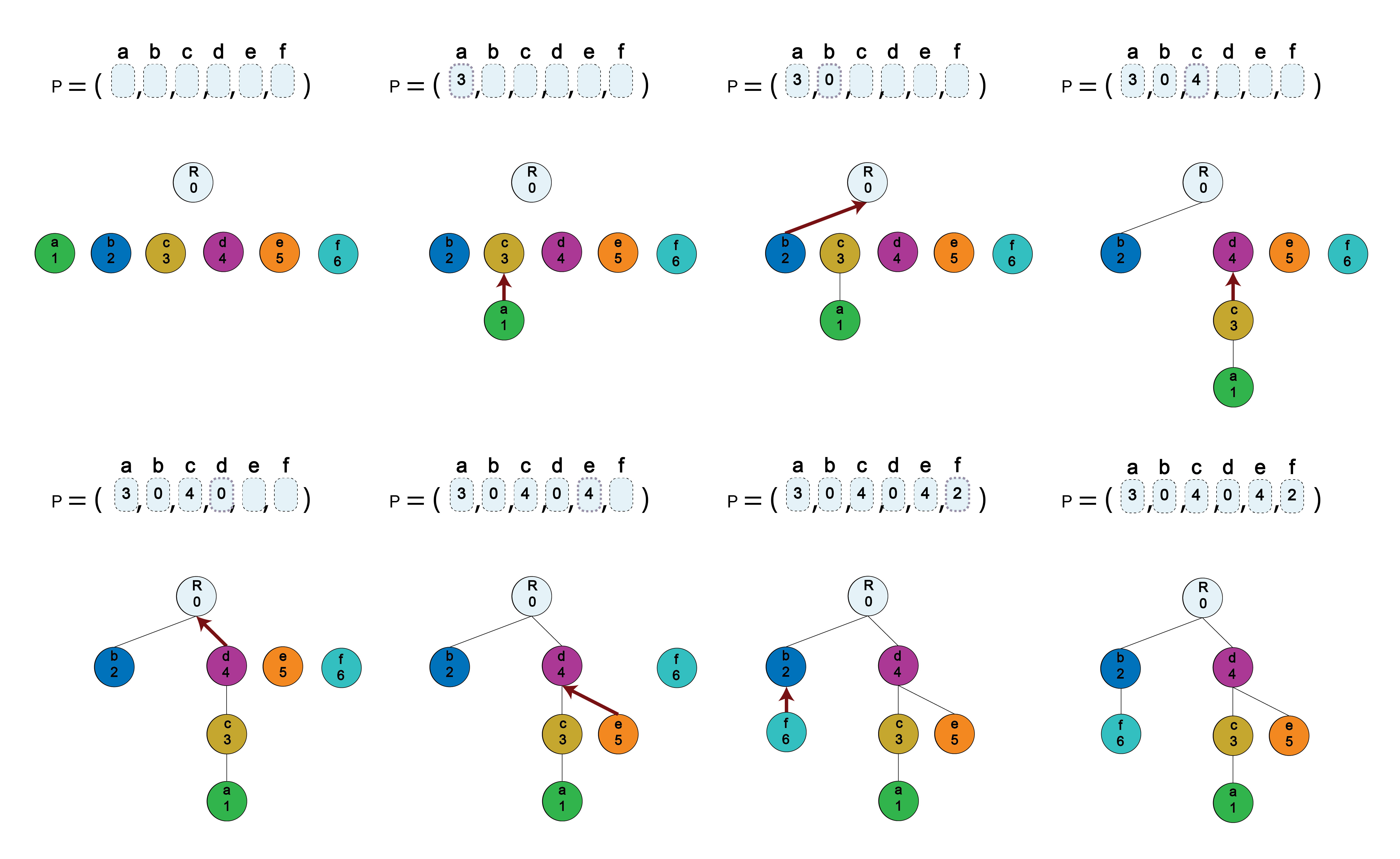}
  \caption{Proceeding Perturbation of O-Tree Method
  \label{fig:o_tree_proc}}
\end{figure*}

The Proceeding Perturbation method converts a Standard O-Tree $T_s = \{ D_0, D_1,.. D_n\}$ into various tree structures via input Permutation Parameters $P = \{p_1, p_2,... p_n\}$ where $p_i \in \{0,1,..,n\} $. To produce alternative floorplanning placements, we perturb the O-tree in $n$ steps, where in each step $i$, $D_{(p_i)}$ becomes the parent of $D_i$. In other words, in each step, the \textit{active node} proceeds the \textit{target node} which holds the index of the corresponding Permutation Parameter. However, the perturbation is discarded if $p_i = i$; or $D_i$ is to proceed a target node that has been defined as it's child or descendant in previous steps. Figure~\ref{fig:o_tree_proc} shows the permutation of an O-Tree via a given $P$ and it's tree representation using the Proceeding method perturbation.

 \subsection{Ascend/Descend Perturbation Method of B*-Trees }\label{sec:adper}
\begin{figure*}
  \centering
  \includegraphics[width=1\columnwidth]{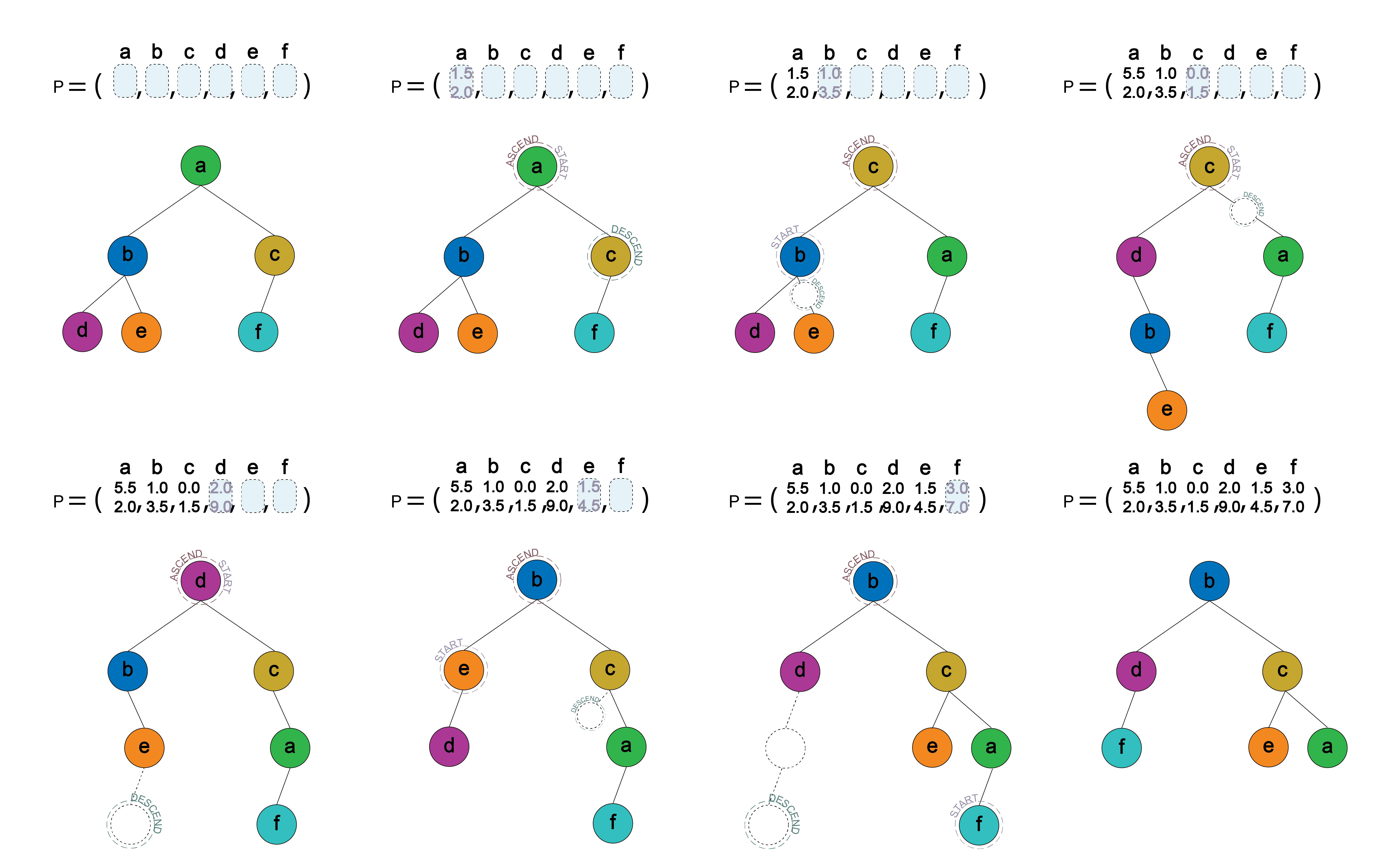}
  \caption{Ascend/Descend of B*-Tree Method
  \label{fig:b_tree_AscDes}}
\end{figure*}

Similar to the Proceeding Perturbation method, the Ascend/Descend method takes place in $n$ steps, wherein each step the active node is to be relocated or swapped with another node. We define the target node as the new position of the active node. Starting with Standard Binary Tree, the permutation parameters of the Ascend/Descend method are defined as $P = \{(p_{1_U}, p_{1_D}),(p_{2_U}, p_{2_D}),..., (p_{n_U}, p_{n_D}) \}$, where $p_{i_U},p_{i_D} \in \{0,0.5,1,1.5,..,2n\}$ . $p_{i_U}$ indicates the maximum levels the node moves up (ascend) in a tree, and $p_{i_U}$ indicates the maximum number of steps the node is to move down (descend) in the binary tree in a breadth-first search (BFS) format. Half-step traversals are also defined to allow the insertion of a node between two levels. After locating the end position of the node relocation, if the end position is occupied by another node, swapping takes place. If no other node occupies the target position, the node creates an internal (a position between two nodes) or an external branch within the tree. We follow the similar procedures for deleting, inserting, and switching block function nodes explained in \cite{Chang_Chang_Wu_Wu_2000}'s study. Additional details of the different operations of the Ascend/Descend method can be found in Table \ref{tab:operations}. Figure \ref{fig:b_tree_AscDes} illustrates the permutation steps of the Ascend/Descend perturbation method.

\begin{table*}
      \caption{Ascend/Descend scenarios and operations.} \label{tab:operations}
\begingroup
\setlength{\tabcolsep}{10pt} 
\renewcommand{\arraystretch}{1.3}
    \begin{tabular}{p{1.5cm} p{4cm}  p{6cm} }
    \hline
    {\small\textbf{Function ID}} & {\small\textbf{Scenario}} & {\small\textbf{Operation}} \\ \hline
 {\textbf{SU}} & {\small{ Step up}} & {\small{Target node is equal with target node parent $T = T_p$. }} \\
 
   {\textbf{HSU}} &{\small{Half step up}} & {\small{Target node is inserted between target node parent and target node.}} \\
  
   {\textbf{SR}} &{\small{Up step count is larger than node level $N_up > N_l$}} & {\small{Target node is equal to root node $T = D_0$.}} \\
  
   {\textbf{HSN}} &{\small{Up step is 0.5 $N_up = 0.5$ }} & {\small{Active nodes equal to target node and graph stays unchanged.}} \\
  
   {\textbf{SD}} &{\small{Step down}} & {\small{Target node is updated with next graph member in BSF format.}} \\
  
    {\textbf{HSD}} &{\small{Half step down}} & {\small{Target node is placed between Ndown ceiling and Ndown ceiling parent.}} \\
   
     {\textbf{SW}} &{\small{Target node points to a graph member}} & {\small{Active node and target node switch places.}} \\
    
    {\textbf{IN}} &{\small{Target node is not graph member and is between two graph members}} & {\small{Active node is deleted and inserted between two graph member nodes.}} \\
   
    {\textbf{DN}} &{\small{Node is deleted and has no child}} & {\small{Parent node pointer to deleted node is set to NULL.}} \\
   
    {\textbf{DL}} &{\small{Node is deleted and has left child.}} & {\small{Node is switched with left child and then deleted.}} \\
 
    {\textbf{DR}} &{\small{Node is deleted and has right child only}} & {\small{Nodes is switched with right child and then deleted.}} \\
   
    {\textbf{SP}} &{\small{Target node has parent that is not graph member}} & {\small{Target node is switched with previous step until target node has parent that is a graph members.}} \\
   \hline
    \end{tabular}
    \endgroup
    
\end{table*}

 \subsection{Available Nodes of B*-Trees}\label{sec:avper}
\begin{figure*}
  \centering
  \includegraphics[width=1\columnwidth]{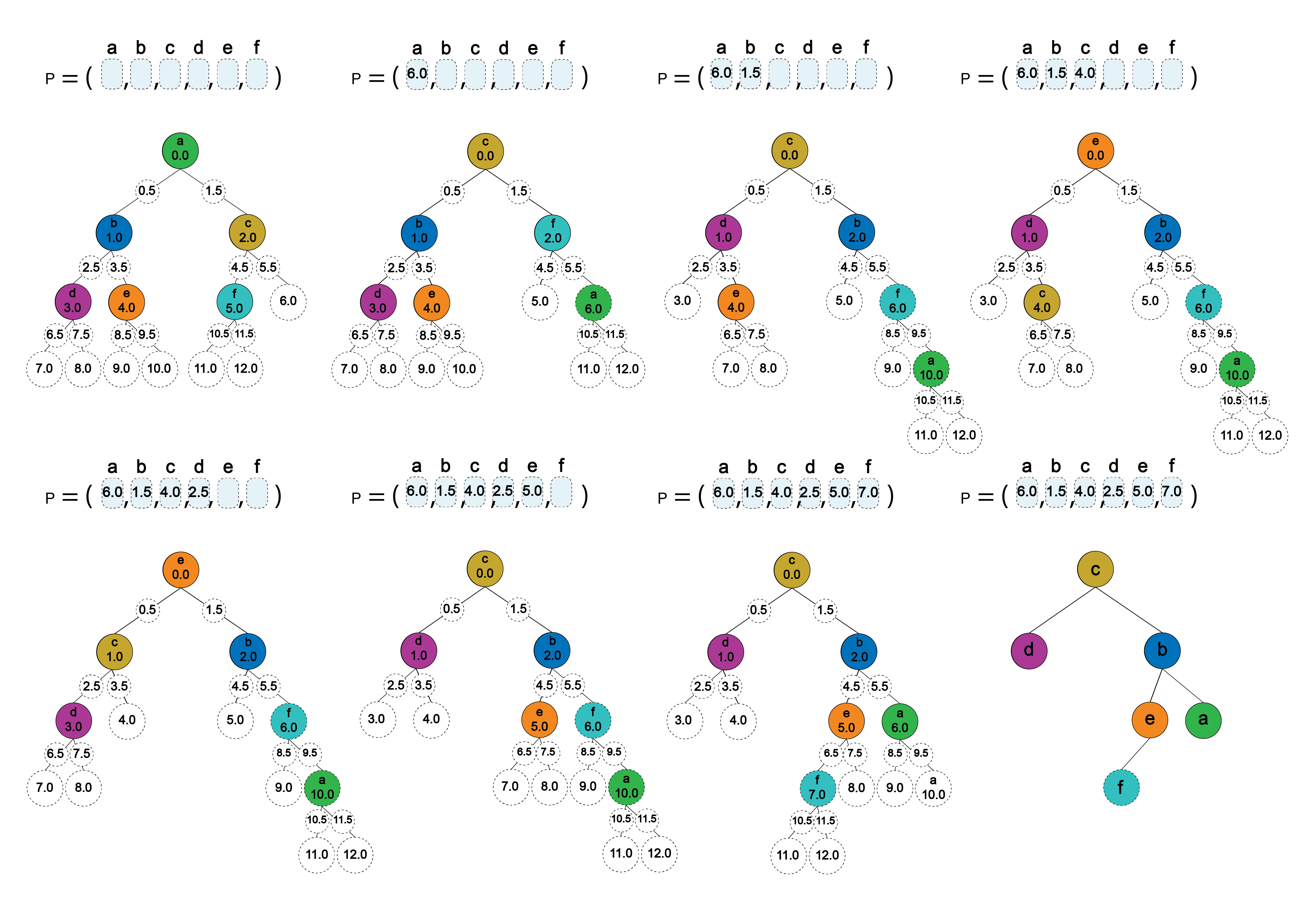}
  \caption{Available Nodes of B*-Tree Method
  \label{fig:b_tree_aval}}
\end{figure*}

Similar to the Ascend/Descend method, the Available Nodes method aims to relocate nodes of a input Standard binary-tree $T_s = \{ D_1, D_2, D_3,.. D_n\}$ in $n$ steps, each time relocating one active node. The Permutation Parameters for this method is defined as $P = \{p_1, p_2,... p_n\}$, where $p_i \in \{0, 0.5, 1, 1.5,..., {2n}\}$. For each binary-tree $T$ with $n$ nodes, there are $4n$ possible target locations in which a node can be relocated resulting in a creation of either an internal node, external node, or a node swap. We call these potential target node locations \textit{available nodes} and represent them in BSF order in $T_{Av} = \{D^\prime_0, D^\prime_{0.5}, D^\prime_1,... D^\prime_{2n}\}$. In each step, $D_i$ will be relocated to $D^\prime_{p_i}$ using the same operations used in the Ascend/Descend method (Table \ref{tab:operations}), and $T_{Av}$ will be updated given the new tree structure. In contrast to the Ascend/ Descent method where the solution space holds a $2n^n + (4n)^n$ complexity, the Available nodes method results in $(4n)^n$ complexity achieving faster results. Figure \ref{fig:b_tree_aval} illustrates the process of the Available Nodes method.

\subsection{Extended B*Tree}\label{extendSection}
While both O*Tree and B*Tree are tree representations of admissible non-slicing floorplan layouts, they can also be extended as slicing methodologies for a given spatial boundary. Such an approach would facilitate the layout planning in scenarios where all the given boundary is expected to be used without leaving empty space. Converting non-slicing representations to slicing representations can extend the potential applications of our work to indoor furniture layout. Given an input boundary space, the system generates allocated areas for each furniture entity, while maintaining object-to-object topological requirements between objects.

\begin{figure*}[b]
  \centering
  \includegraphics[width=1\columnwidth]{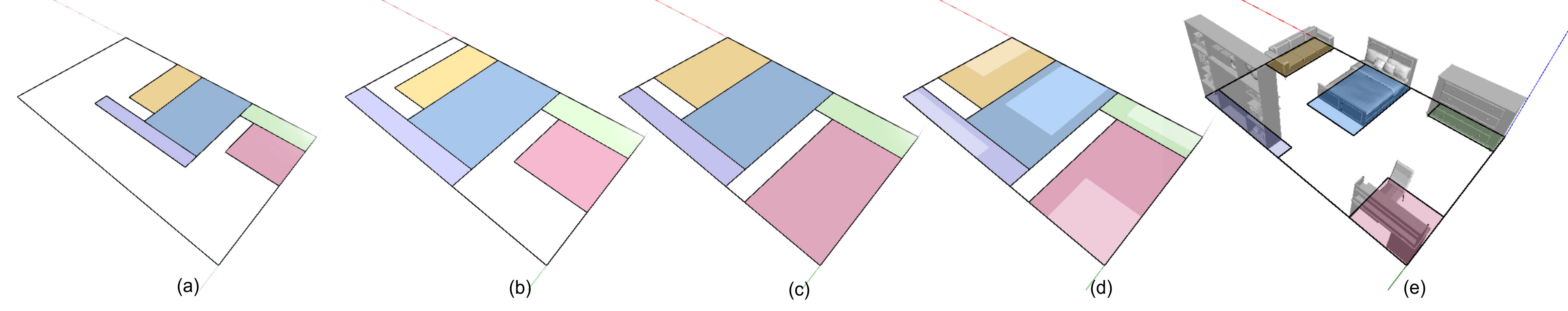}
  \caption{Extended B*Tree process: (a) generate B*Tree layout (b) scale the generated layout (c) extend to the closest boundary edge (d) allocate furniture location based on heuristic constraints (e) add furniture objects to location.
  \label{fig:exBtree_pr}}
\end{figure*}

\begin{figure*}
  \centering
  \includegraphics[width=1\columnwidth]{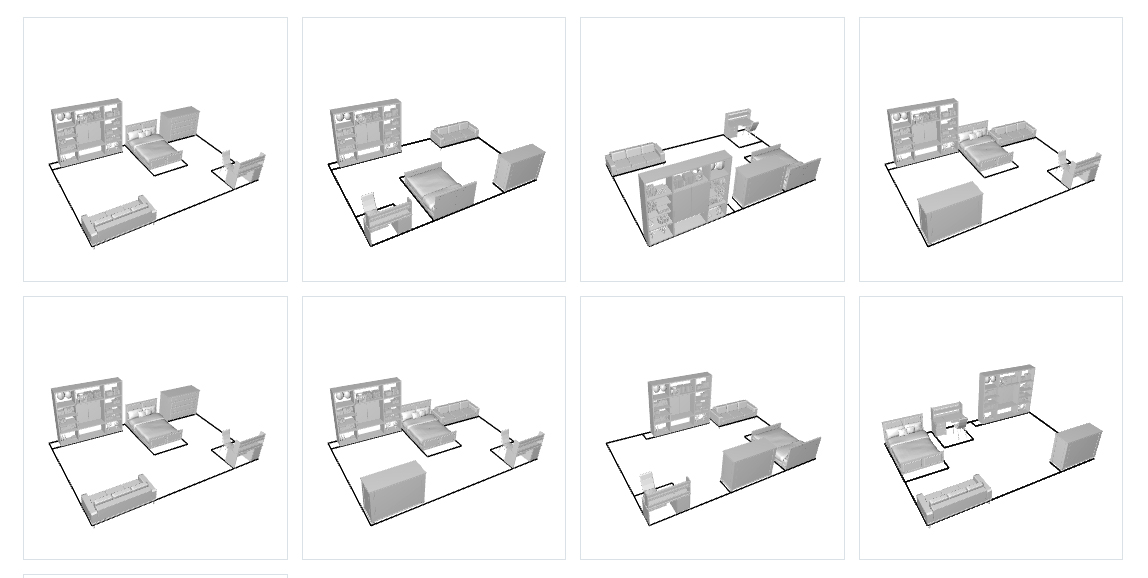}
  \caption{Examples of indoor furniture generation via the Extended B*Tree algorithm
  \label{fig:exBtree_ex}}
\end{figure*}

We implement our extended B*Tree method in four main steps: (i) generate B*Tree layout; (ii) scale the generated layout height and width to fit the input bounding area. We call the new blocks, “extended blocks”. If the ratio of the height or width is bigger than one, we penalize the B*Tree solution as the function blocks cannot fit within the bounding space without overlapping each other; (iii) We extend each extended block to the closest boundary edge, if not blocked by another extended block. This would allow the extended blocks to fill all available spaces within the boundary space. (iv) Finally, for non-flexible block functions such as furniture, where the dimension of the block should be constant, we allocate a section of the new extended blocks as the final location of the block while considering block constraints such as direction and edge requirements. This step does not execute for blocks that can have flexible dimensions. Figure \ref{fig:exBtree_pr} illustrates the process and Figure \ref{fig:exBtree_ex} shows examples of indoor furniture generation using the Extended B*Tree algorithm.

\section{Evaluation and Search}
The resulting floorplan can be then analyzed by various customized fitness functions corresponding to the project requirements. In this module, performance evaluations and simulations are incorporated to evaluate the fitness of the floorplan. Other than the geometrical and topological analysis functions provided by GenFloor, users can integrate third-party libraries and simulators to analyze the generated layout. GenFloor provides basic evaluation functions that can be used as objectives or constraints of the search system. The number of adjacency goals achieved, bounding boxes, and closest distances between blocks can be calculated using native functions available in the GenFloor package.

Adjacency evaluators allow users to analyze how the generated floorplan achieves topological and adjacency goals. The evaluator (1) calculates all adjacent blocks of each Spatial Block in a given list (resulted adjacency); (2) compares the resulted adjacencies and target adjacencies (goals) informing the user whether each resulted adjacency is present in the target adjacency list or not (adjacency check); and (3) outputs the number of achieved adjacencies in all the Spatial Blocks (adjacency achieved count). Other evaluators that are incorporated in the system include bounding area calculation (returns smallest enclosed rectangle, or the minimum bounding rectangle for the generated floorplan), closest distance (returns the closest distance of Spatial Block to a single or multiple Spatial Block and bounding curves test (Calculates whether Spatial Blocks of a generated floorplan fall inside a given polygon).

Finally, GenFloor can take advantage of Dynamo's integrated Non-Sorting Genetic Algorithm II (NSGA-2) which would allow the user to define constraints and goals for the floorplanning problem. The solver would find Permutation Parameters that correspond to tree representations of optimal solutions. While GenFloor is tested to work smoothly with Dynamo's NSGA-2 algorithm, the generation module of the system is independent of the search mechanism, and therefore, can be replaced other search mechanisms such as simulated annealing, particle swarm optimization or other evolutionary solvers, etc. Such approaches have not been experimented with within this study, however, given the modular development of our GenFloor library, such integration is feasible.

\section{BIM Integration and Visualization}\label{sec:viz}
\begin{figure}
  \includegraphics[width=1\columnwidth]{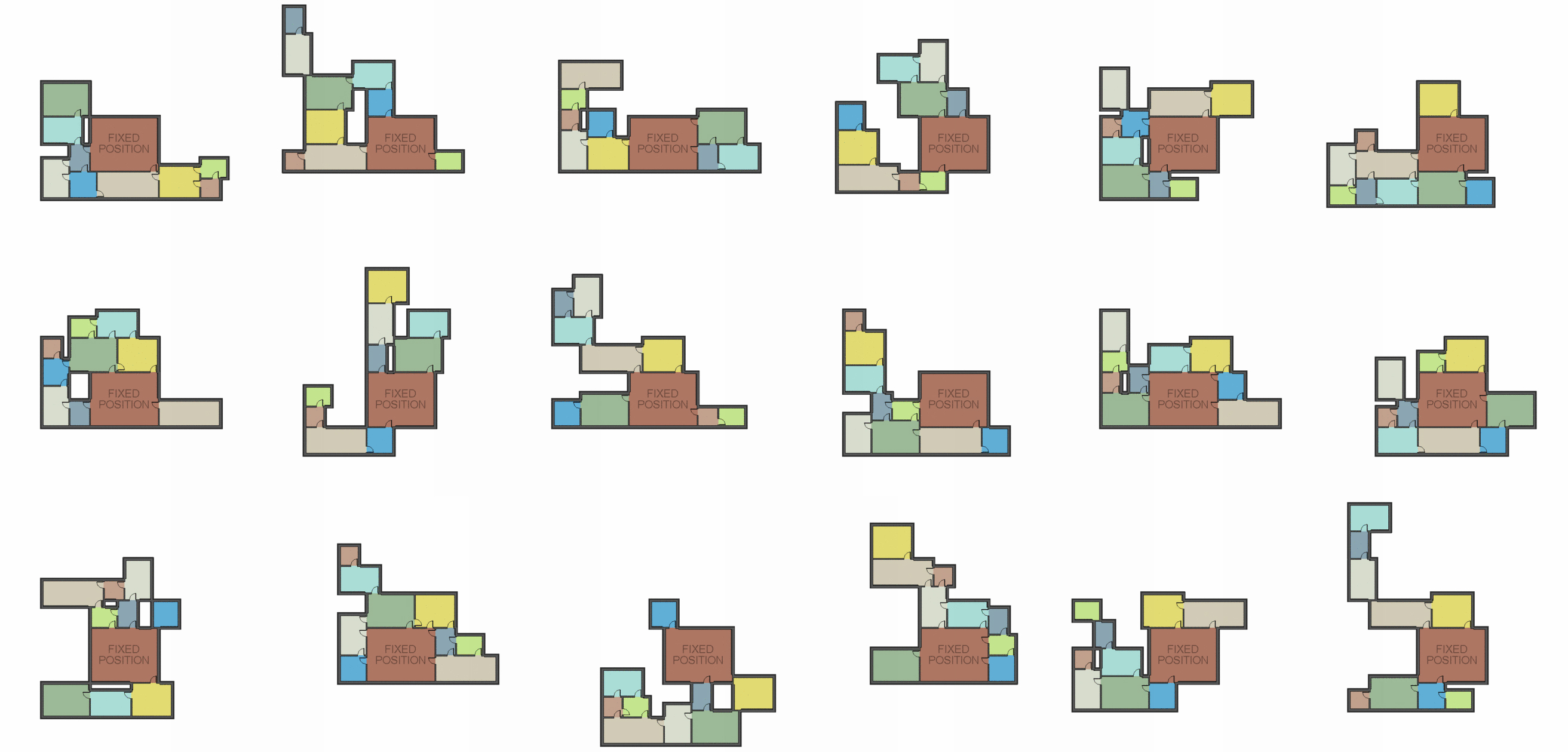}
  \caption{Examples of different perturbations of a floorplan using GenFloor in Revit. The user can generate BIM models including walls, doors and rooms of the populated GenFloor floorplans using built-in components}
  \label{fig:revit}
\end{figure}

Generated floorplans can be directly converted to BIM models using incorporated functions available in the GenFloor library. Users can generate walls, doors, and rooms of the populated GenFloor floorplans in Revit. Spatial Blocks are converted to two sets of non-overlapping walls representing the interior and exterior walls. Moreover, meta-data such as room names, spaces IDs, etc. can be transferred to Revit for additional documentation purposes. A procedural door placement system is also available for users to further automate the process of floorplan generation. Figure \ref{fig:revit} illustrates an example set of results of BIM models, generated as an output of the GenFloor system.

\begin{figure}[b]
  \includegraphics[width=1\columnwidth]{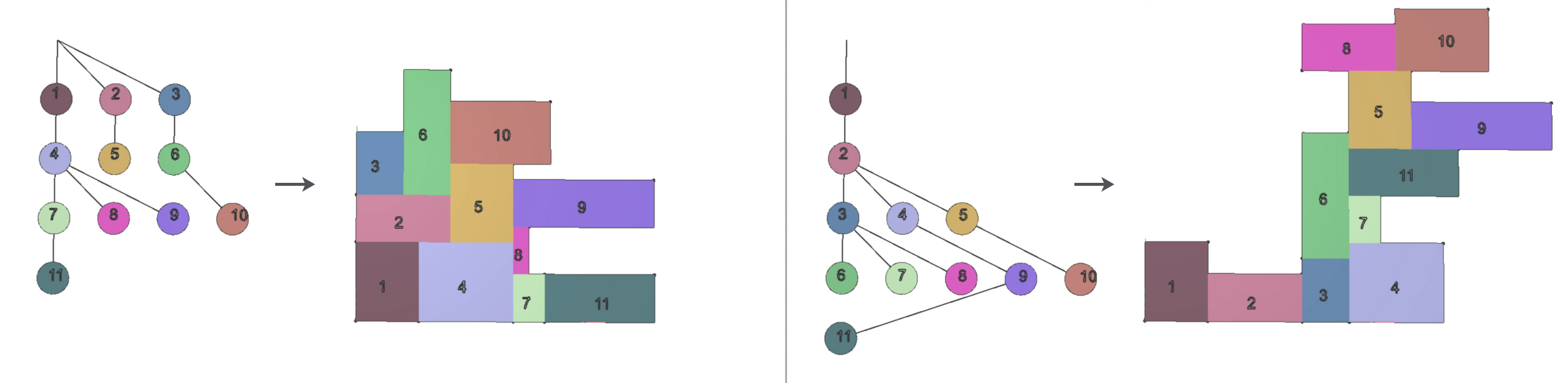}
  \caption{Examples of an O-Tree graph visualization (left) corresponding to the generated floorplan (right) using the visualization functions.}
  \label{fig:treeViz}
\end{figure}

Furthermore, users can visualize the corresponding adjacency graphs of the generated layout with color codes determining which of its topological and adjacency goals it has been achieved and which has not. Adjacency graphs or bubble diagrams are a common method of representing topological relationships and requirements in a space layout problem. The corresponding tree structure (O-Tree or B*Tree) of the generated space layout can also be generated by the user to allow additional analysis of the system workflow. Examples of such an approach can be seen in Figure \ref{fig:treeViz}.

\section{Case Studies}
 
In performance-based generative design systems, the general goal is to find optimal solutions that address the constraints and goals defined by the user. Yet in applications such as space layout planning, automation can serve in various stages of the design process with different objectives. In design scenarios where goals and constraints are finalized, programmable, and clearly defined, the user is simply seeking the best solution (or a limited number of solutions) which achieves optimal performance. In such settings, an ideal performance-based generative design system is able to generate and converge towards the global optima of the solution landscape without getting trapped in local optimas.

The other scenario is when all performance metrics are not programmable by the designer or have not yet finalized the performance metrics for its design. Programming evaluators often require time and expertise, and not always can programmable due to the qualitative nature which they may hold. In such cases, the designer may be yet flexible with the defined metrics and would want to explore different design strategies. In such a scenario the user is looking for more diverse solutions. Instead of looking for one final answer, the designer is eager to explore a diverse set of solutions, all addressing the goals and constraints to a good level of extent. Diverse design options sampled from different parts of the solution landscape can provide the user with multiple design directions or initiate a feedback loop so a redefinition of the design problem takes place.

We, therefore, conduct two preliminary experiments to address both scenarios explained above.  In the first experiment, we measure the system’s ability to find a known optimal solution, given that that goals and constraints are finalized and defined. In the second experiment, we observe how the system can generate diverse floorplans while addressing given a constant design problem. From a search perspective, we intend to measure the ability of the system to converge to the global optimum in the first experiment, while in the second experiment assessing its ability to gather all potential solutions for the user often corresponding to local optimums of the solution landscape. We further elaborate on the experiments below.

 \subsection{Optimal Solution Search}

The goal of this experiment is to evaluate whether our system can find a deterministic optimal solution for a given space layout problem. To do so, we define a pre-designed layout as an optimal design. We then use the topological properties of a pre-defined design (optimal design) as goals and assess if the system is able to find a solution similar both in topological performance and geometrical layout. Such an approach would allow us to observe whether the system would (i) converge to a deterministic optimal solution in different runs, (ii) compare the speed and memory allocation for various permutation methods, and finally (iii) understand how various properties of the evolutionary solver such as population size, generation count and mutation rate impacts the ability of our system to find the optimal design.

\begin{figure*}
  \centering
  \includegraphics[width=1\columnwidth]{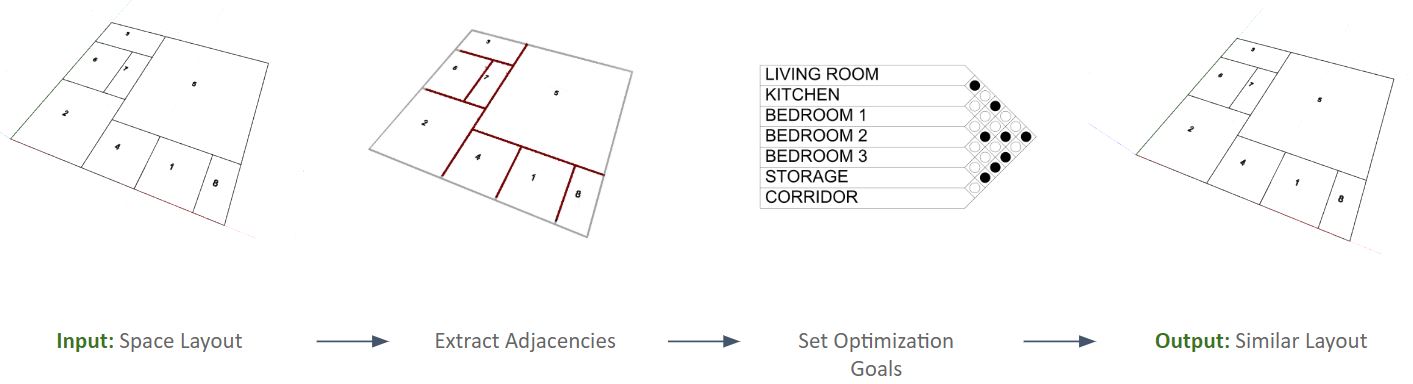}
  \caption{Optimal solution search experiment workflow.
  \label{fig:opt_workflow}}
\end{figure*}

Figure \ref{fig:opt_workflow} shows the general workflow of the experiment. We use a residential floorplan as a pre-designed optimal layout with 8 spatial blocks, each corresponding to a spatial function. Next, by extracting the topological properties of each block, we produce the adjacency matrix of the floorplan. This would serve as input to the GenFloor system- as an optimization goal- alongside the geometrical properties of the corresponding blocks. We then examine whether the system would generate the ideal output: a floorplan with the maximum adjacency goals achieved and a spatial layout similar to the layout defined as the optimal design. 

We perform the experiment with three different settings of the GA solver. In the first setting, we initiate a limited population size $(p = 40)$ with limited generation runs $(g = 5)$. The second setting resembles a larger population with the same limited generation run $(p=100, g=5)$, while the third setting consists of a larger population with a larger number of generations. The cross-over rate for all runs is 20\% and the mutation rate is set at 10\%. Moreover, we define adjacency goals in three levels. The $L_1$ contains all the adjacency requirements $(A_{Count} = 28)$ extracted from the pre-designed layout, the $L_2$ resembles high priority adjacencies from a designers perspective $(A_{Count} = 20)$. In $L_3$, if we require $S_A$ to be adjacent to $S_B$, while $S_B$ is adjacent to $S_C$, then $S_C$ adjacency to $S_A$ will not be assigned as a goal. In such context, the number of adjacency required where decreased to $(A_{Count} = 16)$. 

Table \ref{tab:optimal} shows the results for our optimal solution search experiment. In $L_1$ where all extracted adjacencies are present, the system is not able to find the optimal pre-designed solution with the various settings and does not meet all the adjacency goals. The maximum score is 0.86 resembling an over-constrained problem for the system. However, when the adjacency requirements are relaxed to a lower amount $L_1$, the system is able to generate the exact optimal layout. In such a context, $(p=40, g=4)$ starts with lower scores than other levels, but with increasing the gene population and generation runs, the system is able to find the optimal solution. Moreover, In the event where adjacency goals are minimum (a = 16), we observe better adjacency scores in $S_{(40,5)}$ and $(p=100, g=5)$ than other adjacency levels. While all adjacency goals are achieved in $(p=100, g=15)$ the output solution is not similar to our pre-designed optimal layout.

 \begin{table*}
      \caption{Optimal Solutions Search.} \label{tab:optimal}
\begingroup
\setlength{\tabcolsep}{10pt} 
\renewcommand{\arraystretch}{1.3}
    \begin{tabular}{ p{2cm} p{2.2cm} p{2cm} p{2cm} p{2cm} p{2cm}}
    \hline
{\small\textbf{Adjacency Required}}&{\small\textbf{Adjacency Count $(A_{Count})$}} &  {\small\textbf{Population Size $(p)$}} & {\small\textbf{Generations $(g)$}} & {\small\textbf{Max Adjacency Achieved}} & {\small\textbf{Score}} \\ \hline
    \hline
    \multirow{3}{*}{\includegraphics[width=1.9cm, height=1.5cm]{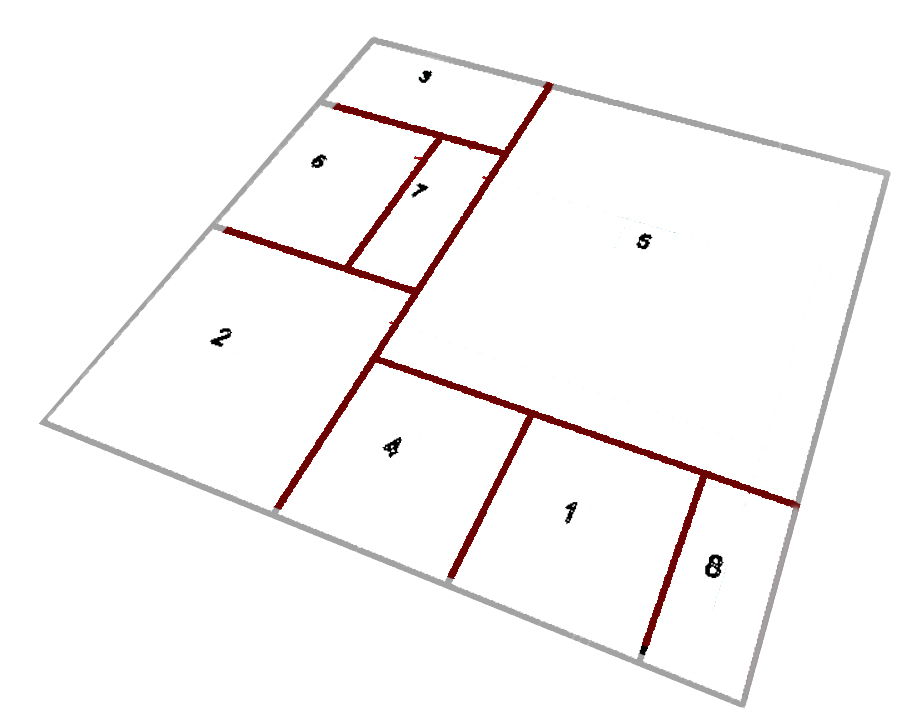}}&{\small{28}} & {\small{40}} & {\small{4}} & {\small{20}} & {\small{0.1}}   \\
 {}&{\small{28}} & {\small{100}} & {\small{5}} & {\small{24}} & {\small{0.79}} \\
{}&{\small{28}} & {\small{100}} & {\small{15}} & {\small{24}} & {\small{0.86}}   \\

    \hline
 \multirow{3}{*}{\includegraphics[width=1.9cm, height=1.5cm]{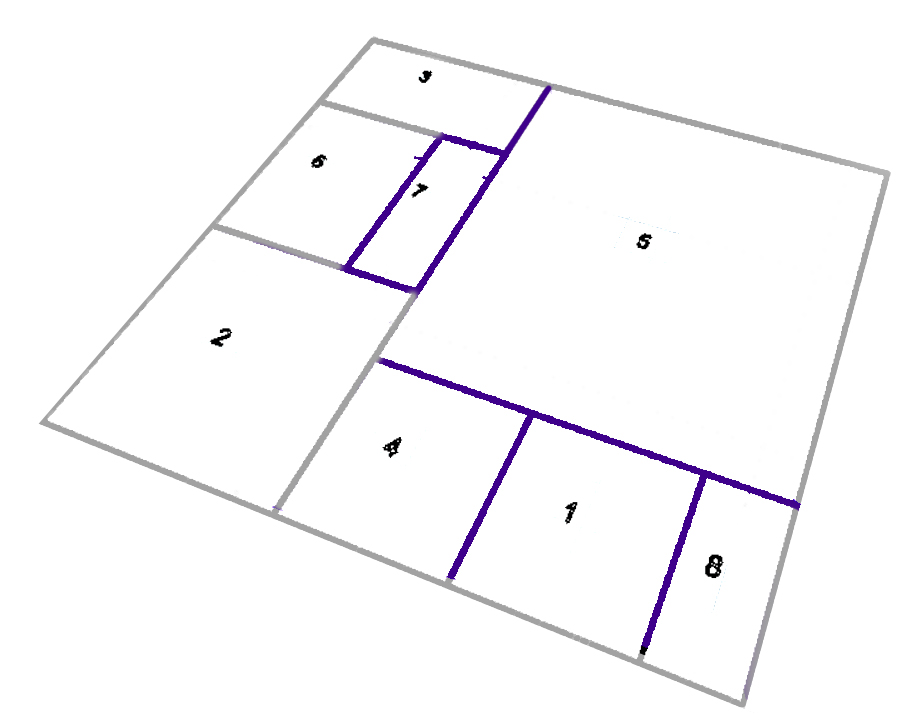}}&{\small{20}} & {\small{40}} & {\small{4}} & {\small{14}} & {\small{0.70}} \\
 {}&{\small{20}} & {\small{100}} & {\small{5}} & {\small{16}} & {\small{0.80}} \\
{}&{\small{20}} & {\small{100}} & {\small{15}} & {\small{20}} & {\small\textbf{1.00}} \\

    \hline
\multirow{3}{*}{\includegraphics[width=1.9cm, height=1.5cm]{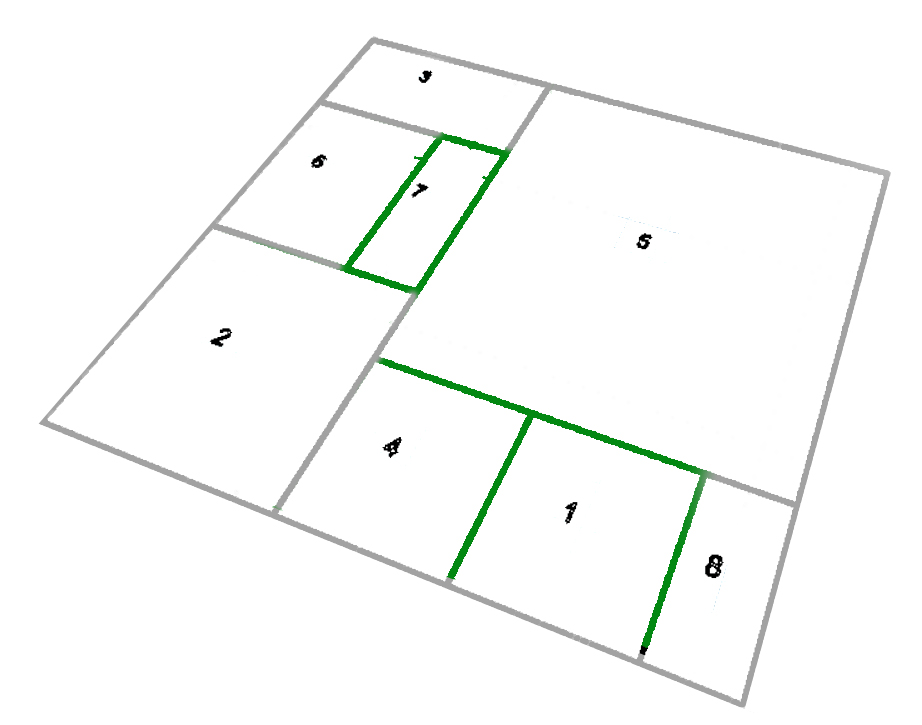}}& {\small{16}} & {\small{40}} & {\small{4}} & {\small{12}} & {\small{0.75}} \\
 {}&{\small{16}} & {\small{100}} & {\small{5}} & {\small{14}} & {\small{0.88}} \\
{}&{\small{16}} & {\small{100}} & {\small{15}} & {\small{16}} & {\small{1.00}} \\

    \hline
    \end{tabular}
    \endgroup
    
\end{table*}

  \subsection{Emerging Layout Search}
\begin{figure*}
  \centering
  \includegraphics[width=1\columnwidth]{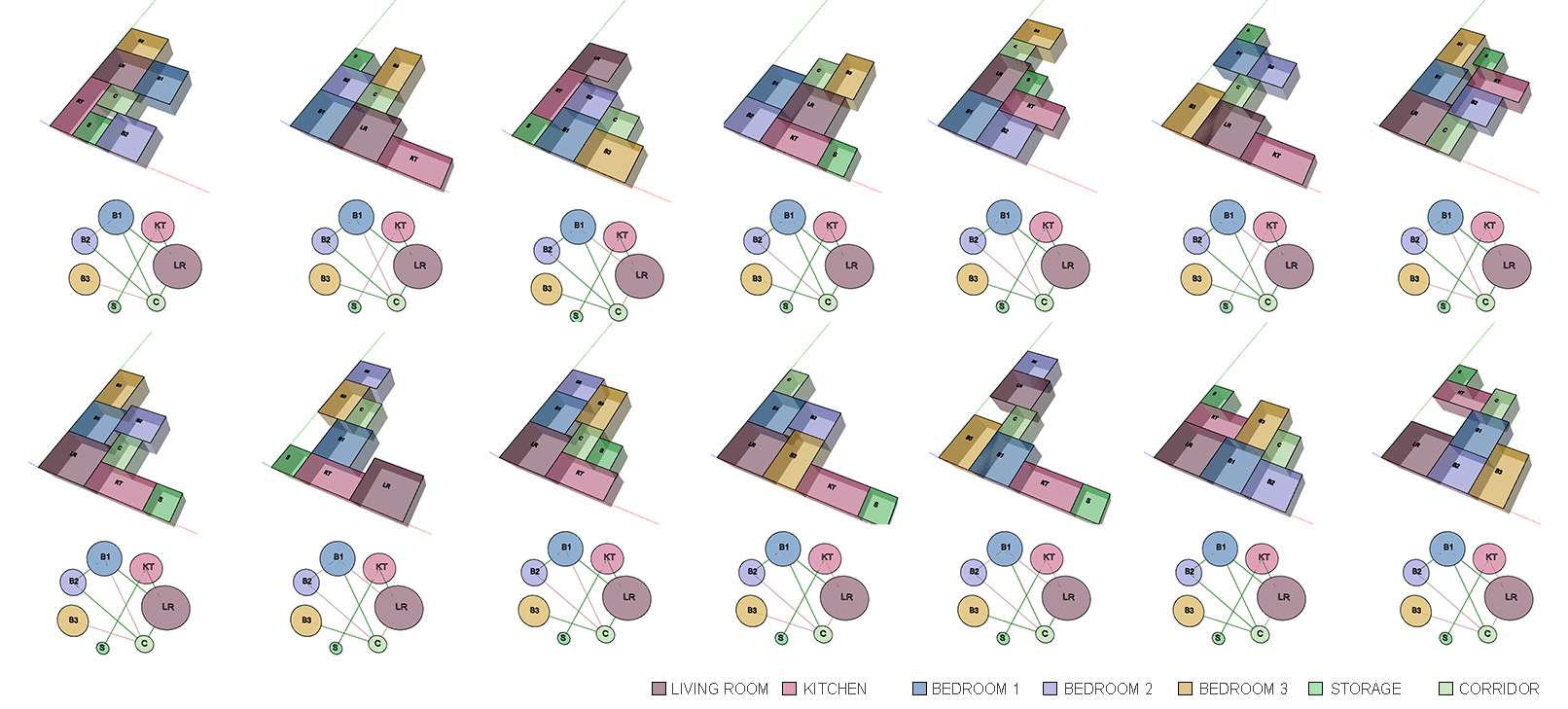}
  \caption{Emerging Layout Search and corresponding bubble diagram visualization generated by GenFloor.
  \label{fig:emerging}}
\end{figure*}

In this experiment, we intend to illustrate our system's ability to generate diverse results given constant geometrical and topological goals and constraints. In addition to the permutation parameters, we add a binary rotation parameter allowing the Spatial Block to switch its width and height. We use the same residential problem in the previous subsection with $L_1$ adjacency requirements where all spaces have at least one adjacency requirement, with the corridor having 4 adjacency goals. However, instead of generating solutions that address all adjacency requirements $(A_{Count} = 28)$, we aim to visualize a broader set of solutions that maintain at least $(A_{Count} = 20)$. Figure \ref{fig:emerging} illustrates the results from a $(p=100, g=15)$ generation and their corresponding bubble diagram visualization explained in Sec \ref{sec:viz}. The results show how GenFloor is able to generate a diverse set of solutions for the defined space layout problem.

 \section{Discussions}
 
 
One of the main drawbacks of using non-slicing floorplanning representation methods in floorplan generation is their lack of intuitiveness between the encoding mechanism and the output floorplanning representation. In such context, while GenFloor aims to provide additional visualization modules for generative designers to view the resulted tree structures of the input array strings, it still falls short of providing a clear and straightforward connection between the input parameters and output results for amateur users. In general, integrating slicing methods with optimization solvers is more straight-forward as control parameters can simply correspond with the $H, V$ parameters. These parameters can be explicit without requiring any post-processing conversation to floorplan representation. Such property does not always hold in non-slicing methods, as encoding methodologies and perturbation operators prevent straight-forward representations of the control parameters.

When comparing some of the limitations of the proposed permutation methods, we should highlight that the O-Tree Proceeding method comes with a larger number of discarding scenarios when compared to the proposed B*Tree permutation methods. Such property may be misleading for an evolutionary search mechanism, as solutions of neighboring parameters may need to be discarded because they don't hold the requirements of permutation parameters, potentially disrupting the ability to converge to optimal solutions. The Ascend/Descend method comes with the limitation of potentially generating similar solutions given different parameters. For example, if the active node is located on the third level of the tree, any $p_{i_U} > 2$ would give the same ascending target node. The Available Nodes method is probably the most efficient permutation model when compared to other methods. However, it lacks an intuitive correspondence to the floorplanning representation. Since the traversal takes place using a BSF approach, understanding the relationship between the permutation parameters and the resulting floorplan can be a challenge for a user. Whereas the Ascend/Descend and the Proceeding methods come with much more intuitive permutation parameters, allowing the user to have potentially more direct control over the floorplanning if desired.

It is important to note, all representations introduced in this paper provide an initial general layout for spatial elements. Additional operators can be implemented to transform the rectangular solutions into more complex geometries, or allocated circulations space in applications such as architecture and facility layout planning. Such an approach would result in more satisfactory results. In scenarios where computationally expensive evaluations take place, such as physical-based simulations (eg. building energy simulations, daylighting simulations, etc.), a limited solution space would be beneficial for the search. Such an approach would also benefit workflows with limited computational resources, such as mobile applications or real-time gaming.

\section{Conclusion}

In this paper, we introduce an end-to-end interactive generative floorplanning design system, GenFloor, which takes geometrical, topological, and performance goals and constraints as input and provides optimized spatial design solutions as output. By proposing three novel permutation methodologies for binary and non-binary trees, and integrating them with genetic algorithms with defined performance evaluators, we are able to generate and optimize Manhattan-based floorplans for various design applications. Our system is developed as a library package for Autodesk Dynamo, a visual programming platform, allowing users to incorporate the generated floorplans with conventional CAD software and third-party performance evaluators for optimization purposes.

Future work on the development of this system can fall into three main categories. First, extending the floorplanning generation to non-Manhattan-based design solutions. This can be explored when integrating physical-based simulations into the current workflow. In such an approach GenFloor can generate initial positioning for the non-rectangular blocks, and then the system can take advantage of mass-spring simulations to configure the geometrical conflicts and resemble a balanced layout. Second, we intend to increase the speed of our system by engineering the current development code with more efficient functions and run-time procedures. The current version released does not resemble the potential speed in which tree representations can deliver in floorplanning optimization.  Initially, additional interactivity in post-modeling generated floorplanning can be added to the current system. While our integration with Revit’s BIM modeling system does allow parametric modification of our generated layouts, but slight re-arrangements of the outputted layouts are not achieved smoothly in the current workflow. Such an approach would enable users to easily implement slight modifications to the generated solutions while obtaining instant feedback to control how the modified layout affects the defined performance criteria


\section{Acknowledgements}
This research was initiated and supported by Autodesk's AEC Generative Design Team. We would like to thank their support in Generative Design for Revit (formerly Project Refinery).


\bibliographystyle{cas-model2-names}

\bibliography{cas-refs}


\end{document}